\documentclass[aps,prd,showpacs,twocolumn]{revtex4}

\usepackage{color}
\usepackage{amssymb,amsmath}

\newcommand{\beq}{\begin{equation}}
\newcommand{\eeq}{\end{equation}}

\newcommand{\h}[1]{{}^{\mbox{\,\tiny $\{#1\}\!$}}h}
\newcommand{\pert}[2]{{}^{\mbox{\,\tiny $\{#1\}\!$}}{#2}}
\newcommand{\lblue}{{\bar l}}
\newcommand{\lred}{{\hat l}}
\newcommand{\mblue}{{\bar m}}
\newcommand{\mred}{{\hat m}}
\newcommand{\Source}[7]{{{}^{\mbox{\tiny $({#1})\!$}}S_{#2}^{#3}
{}_{#4}^{#5}{}_{#6}^{#7}}}
\newcommand{\E}[7]{{E_{#1}^{#4}{}_{#2}^{#5}{}_{#3}^{#6}
{}^{}_{#7}}}
\newcommand{\C}[6]{{C_{#1}^{#2}{}_{#3}^{#4}{}_{#5}^{#6}}}

\newcommand{\Zer}{{\Psi}}

\newcommand{\RW}{{\Phi}}

\begin{document}
\title{A complete gauge-invariant formalism for arbitrary  
second-order perturbations of a Schwarzschild black hole}
\author{David Brizuela}
\affiliation{Instituto de Estructura de la Materia, CSIC, Serrano
121, 28006 Madrid, Spain}

\author{Jos\'e M. Mart\'{\i}n-Garc\'{\i}a}
\affiliation{
Laboratoire Univers et Th\'eories, Observatoire de Paris, CNRS,
Univ. Paris Diderot, 5 place Jules Janssen, 92190 Meudon, France, and \\
Institut d'Astrophysique de Paris, Univ. Pierre et Marie Curie,
CNRS, 98bis boulevard Arago, 75014 Paris, France}

\author{Manuel Tiglio}
\affiliation{Center for Scientific Computation and Mathematical Modeling
and\\
Center for Fundamental Physics, Department of Physics, University of
Maryland, College Park, MD, 20742}

\begin{abstract}
Using recently developed efficient symbolic manipulations tools, we
present a general gauge-invariant formalism to study
arbitrary radiative $(l\geq 2)$ second-order perturbations
of a Schwarzschild black hole.
In particular, we construct the second order Zerilli and Regge-Wheeler
equations under the presence of any two first-order modes, reconstruct
the perturbed metric in terms of the master scalars, and compute the
radiated energy at null infinity.

The results of this paper enable systematic studies of generic second
order perturbations of the Schwarzschild spacetime. 
In particular, studies of mode-mode coupling and non-linear effects in
gravitational radiation, the second-order stability of the 
Schwarzschild spacetime, or the geometry of the black hole horizon.


\end{abstract}

\maketitle

\section{Introduction}

Nonlinearities are a characteristic feature of General Relativity.
Classifying the possible types of nonlinearity can help us much
in understanding and interpreting the results of observations or
simulations of gravitational phenomena. For instance, the space-based
gravitational-wave observatory LISA will be able to detect quasinormal
mode coupling and frequency mixing in the ring-down phase of the
collision of two supermassive black holes at distances of 1 Gpc.
Perturbation theory offers a systematic study of nonlinearities,
at least for moderate couplings, and it is hence worth developing
a complete formalism for perturbations in General Relativity.
%

During the last three years we have prepared a combination of ideas
and tools on which such systematic approach to metric perturbation theory
in General Relativity can be based. This includes, among other things,
general explicit formulas for the perturbations of curvature
tensors at any order \cite{Brizuela:2006ne}, a general analysis
of gauge-invariance \cite{Brizuela:2007zza}, and the construction
of specialized and efficient tensor computer algebra tools to handle
the enormous equations of perturbative General Relativity 
\cite{Brizuela:2008ra,xAct}.
For definiteness, and because this is the most standard scenario
in astrophysical applications, we have generally focused on spherical
background spacetimes.

The purpose of this article is to apply that general formalism to a
Schwarzschild background. Linear perturbations of Schwarzschild have
been studied since the pioneering work of Regge and Wheeler
\cite{Regge57}, and later Zerilli \cite{Zerilli70} and Moncrief
\cite{Moncrief:1974vm}, in which they were already able to identify and
describe the two polarizations of gravitational waves, giving two
decoupled wave equations for them.

The first studies of second-order black hole perturbations  were
carried out in the seventies \cite{Tomita74, Tomita76} in order
to study the non-linear stability of the Schwarzschild solution.
In the mid-nineties, motivated by the close-limit approximation
\cite{1994PhRvL..72.3297P}, gauge-invariant second-order generalizations
of the Regge-Wheeler and Zerilli-Moncrief formalisms were developed and
applied to a variety of close limit-type initial data for binary black
holes; see \cite{Gleiser:1998rw,Nicasio:2000ge} and references therein.
More recently,
the relevance of second-order perturbation theory on an extreme
mass-ratio inspiral (EMRI) has been evaluated \cite{Lousto:2009}.
In general, those references have focused on the study of the
self-coupling of a particular first-order mode and how that generates
some second-order mode.



We present here a complete, gauge-invariant Regge-Wheeler-Zerilli-Moncrief
like formalism for {\em arbitrary} $l\ge 2$ first and second-order
perturbations of a Schwarzschild black hole. 
In particular, we derive the general first and second order
Regge-Wheeler and Zerilli master variables and the equations they obey. 
We also reconstruct the perturbed metric in terms of those scalars,
as well as computing the radiated energy at null infinity. 

Most of our results are not only gauge invariant but also covariant;
that is, independent of the background coordinates. This is analogous
to the covariant formalisms for linear perturbation theory in
references \cite{Gerlach79, Gerlach:1980tx} and
\cite{Sarbach:2001qq,Martel:2005ir}. This is a necessary step to,
for example, test the geometry near the horizon or null infinity,
for which Schwarzschild coordinates are inadequate. Other results
are still given in Schwarzschild coordinates, for simplicity.

The organization of this paper is as follows: we start in Section
\ref{sec:GS} by reviewing the formalism of
\cite{Brizuela:2006ne, Brizuela:2007zza}. 
Section \ref{sec:high} presents the high order Regge-Wheeler and Zerilli
equations and Section \ref{sec:power} sketches how we compute 
the radiated energy. Setting up the arena for the second order treatment,
Section \ref{sec:first} rederives in a compact yet complete way a
gauge-invariant version of the Regge-Wheeler-Zerilli first order formalism.
Finally, Section \ref{sec:second} presents our general second order 
Regge-Wheeler and Zerilli equations, their sources, and the computation
of the radiated energy in terms of our second order Regge-Wheeler and
Zerilli functions. 

\section{High-order Gerlach and Sengupta formalism} \label{sec:GS}

\subsection{Background spherical spacetime}

This section briefly summarizes the formalism introduced in
references \cite{Brizuela:2006ne, Brizuela:2007zza} to deal with high-order
perturbations of a spherical spacetime. This formalism can be
regarded as the generalization to higher orders of the
Gerlach-Sengupta linear formalism \cite{Gerlach79, Gerlach:1980tx}.

We start by decomposing the spacetime manifold ${\cal M}$
as a product ${{\cal M}\equiv {\cal M}^2\times S^2}$,
where ${\cal M}^2$ is a two-dimensional Lorentzian manifold and $S^2$ the
two-sphere. We will use Greek letters $(\mu,\nu,\dots)$ for
four-dimensional indices,
capital Latin letters $(A,B,\dots)$ for indices on ${\cal M}^2$ and
lowercase Latin letters $(a, b,\dots)$ for indices on the sphere.
With this notation, and choosing an adapted coordinated system,
a general spherically symmetric background metric is given by
\begin{eqnarray} \label{sphericalgdecomposition}
g_{\mu\nu}(x^D, x^d)dx^\mu dx^\nu &=&
g_{AB}(x^D)dx^A dx^B
\\\nonumber
&&\, + \,r^2(x^D)\, \gamma_{ab}(x^d)dx^a dx^b,
\end{eqnarray}
where $g_{AB}$ is a metric field and $r$ a scalar field, both on the
manifold ${\cal M}^2$, and $\gamma_{ab}$ is the unit metric on the
sphere $S^2$. The respective coordinate systems on the two-dimensional
manifolds can still be freely specified.

We define the following notation for the covariant derivatives associated
with each metric:
\begin{equation}
g_{AB|C}=0,\quad \gamma_{ab:c}=0 \, .
\end{equation}
For future reference, we define the vector $v_A\equiv r_{|A}/r$.

\subsection{Non-linear perturbations}

In perturbation theory one works with a family of spacetimes
$({\cal M}(\varepsilon),\tilde{g}(\varepsilon))$ depending on a
dimensionless parameter $\varepsilon$. The background spacetime is
the member of this family for which $\varepsilon=0$, and it is assumed
to be a known solution of the Einstein equations. Performing a Taylor
expansion around the background metric $g_{\mu \nu}$,
\begin{equation}\label{expansion}
\tilde g_{\mu\nu} (\varepsilon) = g_{\mu\nu}
+\sum_{n=1}^{\infty}\frac{\varepsilon^n}{n!}\pert{n}{h}_{\mu\nu},
\end{equation}
defines the perturbations $\pert{n}{h}_{\mu\nu}$, which 
are tensors on the background manifold. In practice, we will
always work at a given finite maximum order $N$, truncating the series
at that order.

Using the basis of tensor harmonics described in appendix
\ref{tensorharmonics},
we decompose the perturbations of the metric in the following way,
\begin{widetext}
\begin{eqnarray}\label{metricdecomposition}
\h{n}_{\mu\nu} \equiv \sum_{l,m}
\left(
\begin{array}{cc}
\pert{n}{H}{}_l^m\!{}_{AB} \; Z_l^m &
\pert{n}{H}{}_l^m\!{}_A \; Z_l^m{}_b
+\pert{n}{h}{}_l^m\!{}_A \; X_l^m{}_b \\
\pert{n}{H}{}_l^m\!{}_B \; Z_l^m{}_a +\pert{n}{h} {}_l^m\!{}_B \;
X_l^m{}_a & \;\pert{n}{K}_l^m \; r^2\gamma_{ab} Z_l^m +
\pert{n}{G}_l^m \; r^2Z_l^m\!{}_{ab} + \pert{n}{h}_l^m \;
X_l^m\!{}_{ab}
\end{array}
\right).
\end{eqnarray}
\end{widetext}


Perturbations of all relevant curvature tensors in term of metric
perturbations have been given in \cite{Brizuela:2006ne}, in particular
those of the Einstein tensor. The latter have also been decomposed
there in spherical harmonics at second order, giving full equations
of motion for the harmonic coefficients of the metric.

Now we proceed to extract the gauge freedom present in
the previous decomposition (\ref{metricdecomposition}). Per each
perturbative order $n$, four of those ten components are free.

The most natural gauge choice in spherical symmetry is the one introduced
by Regge and Wheeler (RW) \cite{Regge57} for linear perturbations. Here
we use those same conditions at all orders, for $l\ge 2$:
\begin{equation}\label{RW}
\pert{n}{H}{}_l^m\!{}_A=0,\quad\pert{n}{G}_l^m=0,\quad
\pert{n}{h}_l^m=0.
\end{equation}

Alternatively, it is possible to construct explicit gauge-invariant
combinations of those harmonic coefficients, as explained in
\cite{Brizuela:2007zza}. The idea is simply to select a gauge and
define the gauge invariants to be the expression of the general
gauge transformation of the metric functions from an arbitrary gauge
to that selected gauge. Around spherical symmetry, the requirement
that the gauge invariants are local (almost) uniquely selects the
RW gauge as preferred reference gauge, and this is what we will use.

Schematically (see \cite{Brizuela:2007zza} for more details) the
gauge invariants associated the RW gauge are given by
\begin{eqnarray}\label{invariants1}
\pert{n}{{\mathcal K}}_{AB}&=&\pert{n}{H}_{AB}+\left(\frac{r^2}{2}\pert{n}{G}_{|A}-\pert{n}{H}_A\right)_{|B}
\nonumber\\
&+&\left(\frac{r^2}{2}\pert{n}{G}_{|B}-\pert{n}{H}_B\right)_{|A}+\pert{n}{R}_{AB},\\\label{invariants2}
\pert{n}{{\mathcal K}}&=&\pert{n}{K}+2v^A\left(\frac{r^2}{2}\pert{n}{G}_{|A}-\pert{n}{H}_A\right)
\nonumber\\
&+&\frac{l(l+1)}{2}\pert{n}{G}+\pert{n}{R},\\\label{invariants3}
\pert{n}{\kappa}_A&=&\pert{n}{h}_A-\frac{r^2}{2}\left(\frac{\pert{n}{h}}{r^2}\right)_{|A}+\pert{n}{R}_A,
\end{eqnarray}
where the terms $\pert{n}{R}_{AB}$, $\pert{n}{R}_A$ and
$\pert{n}{R}$ are polynomial combinations of the lower order 
harmonic coefficients 
\beq
\{\pert{k}{H}_l^m{}_A,\pert{k}{G}_l^m,\pert{k}{h}_l^m\}, \mbox{ with }  k<n,
\label{eq:kln} 
\eeq
all vanishing in RW gauge, by construction.
Because of the $k<n$ condition, those terms are not present
in linearized perturbations ($n=1$). For second order perturbations
($n=2$), they are quadratic in the first-order quantities 
$\{\pert{1}{H}_l^m{}_A,\pert{1}{G}_l^m,\pert{1}{h}_l^m\}$, and so forth
for higher $n$. 

Once more, the values of
$\pert{n}{{\mathcal K}}_{AB}, \pert{n}{{\mathcal K}}, \pert{n}{\kappa}_A$ 
in any gauge coincide with their values in the RW gauge. Because of this
we can do calculations in the RW gauge, yet still recover the form of
any expression in a generic gauge by making use of the definitions
(\ref{invariants1}-\ref{invariants3}).
We will take advantage of this fact later.

\section{High order Regge-Wheeler and Zerilli equations} \label{sec:high}

{}From now on we shall assume that our background solution is the
Schwarzschild spacetime. The perturbative formalism summarized in the
previous section allows us to work with arbitrary coordinates on the
reduced manifold ${\cal M}^2$.
However we will sometimes use Schwarzschild coordinates $(t, r)$ as
intermediate tools for computing expressions which in the end will
be valid in any asymptotically flat background coordinates. Then we
will use the following shorthands for coordinate derivatives acting
on any object $\omega$:
\begin{equation}
\dot {\omega} \equiv \frac{\partial\omega}{\partial t} ,
\qquad \omega' \equiv \frac{\partial\omega}{\partial r}.
\end{equation}
%

When studying perturbations of the Schwarzschild spacetime, it is
possible to reduce the perturbed Einstein equations to two wave
equations for two {\em master} functions, one of odd/axial parity
and another one even/polar parity, and this is true at all
perturbative orders. These two functions fully describe the
gravitational content of the system, and actually the whole
metric perturbation at order $n$ can be reconstructed from the
master variables at orders $k\le n$.

\subsection{The even/polar parity sector}

We define the $n$th order Zerilli function as the following
combination of polar harmonic coefficients,
\begin{equation}\label{def1}
\pert{n}{\Psi}\equiv\frac{r^4}{3M+\lambda}(v^{B}\pert{n}{\mathcal K}_{AB}-\pert{n}{\mathcal K}_{|A})v^{A}
+r\pert{n}{\mathcal K},
\end{equation}
where $\lambda\equiv\frac{1}{2}(l-1)(l+2)$.
This variable is given in terms of the $n$th order gauge invariants
tied to the Regge-Wheeler gauge. To recover the form of the Zerilli
function in a different gauge, we just need to replace the
invariants by their explicit form in terms of a generic gauge
(\ref{invariants1}-\ref{invariants3}).
Then, the Zerilli variable takes the following form in Schwarzschild
coordinates:
\begin{eqnarray}\label{def2}
\pert{n}{\Psi}&\equiv&\frac{(2 M-r)}{3M+\lambda r}\left\{(2M-r) \pert{n}{H}_{rr}
+r^2 \pert{n}{K}'\right\}\nonumber\\&+&r\pert{n}{K}
+\frac{l (l+1)}{3M+\lambda r} (2 M-r) \pert{n}{H}_r
\nonumber\\
&+&\frac{1}{2} l(l+1) r \pert{n}{G} + \pert{n}{Q}_\Psi \, ,
\end{eqnarray}
where the subindices $t$ and $r$ stand for components of the
different tensors in these coordinates.
In the previous expression $\pert{n}{Q}_\Psi$ collects the terms
$\{\pert{n}{R}_{AB}, \pert{n}{R}_{A},\pert{n}{R}\}$, and hence it is
itself a polynomial in the lower order variables (\ref{eq:kln}).
Notice that in Eq.~(\ref{def2}) the last three terms, including $\pert{n}{Q}_\Psi$,
are zero when imposing the RW gauge.

The Zerilli master function satisfies the following wave equation
\begin{equation}\label{Zerilli}
\pert{n}{\Psi}^{|A}{}_A-V_{\rm Z} \pert{n}{\Psi}=\pert{n}{\cal S}_\Psi \, .
\end{equation}
The source term $\pert{n}{\cal S}_\Psi$ depends on the lower order 
perturbations, while the potential is
\begin{equation}
V_{\rm Z}\equiv\frac{l(l+1)}{r^2}
-\frac{6M}{r^3}\frac{r^2\lambda(\lambda+2)+3M(r-M)}{(r\lambda+3M)^2}.
\end{equation}

In particular, when using the tortoise coordinates $(t,r^*)$ [with 
$r^*=r+2M \ln \left( r/\left( 2M \right)-1 \right)$] the differential
operator takes the following simple form 
\begin{equation}
\pert{n}{\Psi}^{|A}{}_A\equiv
\left(1-\frac{2M}{r}\right)^{-1}\left(-\frac{\partial^2\pert{n}{\Psi}}{\partial t^2}
+\frac{\partial^2\pert{n}{\Psi}}{\partial r^*{}^2}\right).
\end{equation}

Obviously, the first-order source $\pert{1}{S}_\Psi$ is zero. The second order
one can be given as 
\begin{widetext}
\begin{eqnarray}\nonumber
\pert{2}{\cal S}_\Psi&\equiv&\sum_{\lred,\lblue}\sum_{\mred,\mblue}
\frac{l(l+1)}{r}\Source{\epsilon}{\lred}{\mred}{\lblue}{\mblue}{l}{m}
+\frac{r^4}{3M+\lambda r}v^{A}
\left[
\Source{\epsilon}{\lred}{\mred}{\lblue}{\mblue}{l}{m}{}^B{}_{B|A}
-\Source{\epsilon}{\lred}{\mred}{\lblue}{\mblue}{l}{m}{}_{AB}{}^{|B}
-\frac{l(l+1)}{r^2} \Source{\epsilon}{\lred}{\mred}{\lblue}{\mblue}{l}{m}{}_A
\right]\\\nonumber
&-&\frac{r}{4(3M+\lambda r)^2}
\left[
\left(
84M^2+12(l^2+l-5)Mr+2\lambda(l^2+l-4)r^2
\right)\Source{\epsilon}{\lred}{\mred}{\lblue}{\mblue}{l}{m}{}_A{}^A
\right.\\\label{S2zerilli}
&+&\left. 4r^3(12M+2\lambda r)v^{A}v^{B}\Source{\epsilon}{\lred}{\mred}{\lblue}{\mblue}{l}{m}{}_{AB}
\right],
\end{eqnarray}
\end{widetext}
where $(\lred, \mred)$ and $(\lblue, \mblue)$ are a pair of two first-order
modes which contribute to the second-order $(l, m)$ mode. The sources $S$
in the previous expression are explicitly given in reference
\cite{Brizuela:2006ne}. The polarity sign $\epsilon$ is defined as
$\epsilon\equiv(-1)^{(\lred+\lblue-l)}$.

Note finally that the definition of the high-order Zerilli function
(\ref{def1}) is essentially determined up to addition of low-order
gauge-invariant terms. That is, the addition of such low-order terms
would keep the same form of the Zerilli equation (\ref{Zerilli}),
in particular with the same potential $V_Z$, but would change the source
$\pert{n}{S}_\Psi$. The definition given in (\ref{def1}) is just the
simplest possibility, and follows \cite{Garat:1999vr}. We will later
make use of this freedom.


\subsection{Odd/axial parity sector}

The Gerlach and Sengupta (GS) master scalar is defined as the rotational
of the axial invariant vector $\kappa_A/r^2$,
\begin{equation}
\pert{n}{\Pi}\equiv\epsilon^{AB}\left(\frac{\pert{n}{\kappa}_A}{r^2}\right)_{|B}.
\end{equation}
Like the Zerilli function, it is given in terms of some of the RW gauge-invariants
(\ref{invariants1}-\ref{invariants3}). In a generic gauge it takes the form 
\begin{equation}\label{defPi}
\pert{n}{\Pi} = \epsilon^{AB}\left(\frac{\pert{n}{h}_A}{r^2}\right)_{|B} + \pert{n}{Q}_\Pi,
\end{equation}
where $\pert{n}{Q}_\Pi$ is a source term that depends on lower-order
perturbations, and which is zero in the RW gauge.

It obeys the GS master equation,
\begin{equation}
-\left[\frac{1}{2r^2}(r^4\pert{n}{\Pi})^{|A}\right]_{|A}\!+\frac{(l-1)(l+2)}{2}\pert{n}{\Pi}=
\pert{n}{\cal S}_\Pi.
\label{eq:GS}
\end{equation}
The second-order source can be written in terms of the
source of the Einstein equations,
\begin{equation}
\pert{2}{\cal S}_\Pi=i \epsilon^{AB}\sum_{\bar{l},\hat{l}}
\sum_{\bar{m},\hat{m}}
\Source{-\epsilon}{\bar{l}}{\bar{m}}{\hat{l}}{\hat{m}}{l}{m}{}_{A|B}.
\end{equation}
As in the even parity case, the sources $S$ that appear in the previous
expression are explicitly given in reference \cite{Brizuela:2006ne}. 
Equation (\ref{eq:GS}) is a wave equation for the scalar $\pert{n}\Pi$.
It contains all the relevant physical information of the axial sector.
As we will see, all metric components can be algebraically reconstructed
from this scalar.

We will use the above definition for $\pert{n}{\Pi}$ for historical reasons. But,
in fact, a better choice \footnote{In the sense that the differential operator on the left hand side 
 of Eq.~(\ref{eq:RWn}) is the same one of the `standard' Regge-Wheeler and Zerilli equations. } 
 is the rescaled $\pert{n}{\tilde\Pi}\equiv r^3\pert{n}{\Pi}$,
because its evolution equation has no first-order derivatives,
\begin{equation}
\pert{n}{\tilde\Pi}_{|A}{}^{A}- V_{\rm RW} \pert{n}{\tilde\Pi}=-2 r \pert{n}{\cal S}_{\Pi}.
\label{eq:RWn}
\end{equation}
This equation is valid in any spherically symmetric background and the
potential is given by
\begin{equation}
V_{\rm RW}=\frac{l(l+1)}{r^2}-\frac{3}{r^2}(1-g^{AB}r_{|A}r_{|B}).
\end{equation}
For the Schwarzschild spacetime it is
\begin{equation}
V_{\rm RW}=\frac{l(l+1)}{r^2}-\frac{6M}{r^3}, 
\end{equation}
and Eq.~(\ref{eq:RWn}) becomes the standard RW equation.

One of the main advantages of the GS master scalar is that the perturbation
of the metric can be {\em algebraically} reconstructed.
But there are some other variables which obey the same RW equation.
In particular, one that will be very useful for our purposes is the one
introduced by Regge and Wheeler themselves in their seminal paper \cite{Regge57}. Its
gauge-invariant generalization to higher orders takes the following form,
\begin{equation}
\pert{n}{\Phi}\equiv v^A \pert{n}{\kappa_A}.
\end{equation}
When using Schwarzschild coordinates for the background this definition becomes 
\begin{equation}\label{defPhi}
\pert{n}{\Phi} = \frac{2 M-r}{2 r^2}\left[\pert{n}{h}'-2\pert{n}{h_r}
-\frac{2}{r}\pert{n}{h}\right]+\pert{n}{Q}_\Phi,
\end{equation}
where $\pert{n}{Q}_\Phi$ is the standard term that depends on lower order
perturbations and vanishes when particularizing to the RW gauge. There are
practical advantages and disadvantages for each of these definitions for
the master scalar in the odd parity sector, as we will see in Section 
\ref{sec:odd1}.

At linear order, $\pert{1}{\Phi}$ obeys the same equation as $\tilde\Pi$.
But at second order the source term changes,
\begin{equation}\label{RW2eq}
\pert{2}{\Phi}_{|A}{}^{A}- V_{\rm RW} \pert{2}{\Phi}=\pert{2}{\cal S}_{\Phi},
\end{equation}
with
\begin{eqnarray}\label{S2RW}
\pert{2}{\cal S}_{\Phi}&\equiv&\sum_{\lred,\lblue}\sum_{\mred,\mblue}
\frac{2 i}{r^3}(3 M-r)\Source{-\epsilon}{\lred}{\mred}{\lblue}{\mblue}{l}{m}
\nonumber\\
&+&i v^A \left[\Source{-\epsilon}{\lred}{\mred}{\lblue}{\mblue}{l}{m}_{|A}
-2\Source{-\epsilon}{\lred}{\mred}{\lblue}{\mblue}{l}{m}_{A}\right].
\end{eqnarray}

\section{Radiated power} \label{sec:power}

To compute the power radiated to infinity by gravitational waves we
will use the Landau and Lifshitz (LL) formula
\cite{Landau75},
\begin{eqnarray}\label{Landau}
\frac{d{\rm Power}}{d\Omega} &=& \frac{1}{16\pi r^2}\left\{\frac{1}{\sin^2\theta}
\left|\frac{\partial\tilde{g}_{\theta\phi}}{\partial t}\right|^2
\right.\nonumber\\
&+&\left.\frac{1}{4}\left|\frac{\partial \tilde{g}_{\theta\theta}}{\partial t}
-\frac{1}{\sin^2\theta}\frac{\partial \tilde{g}_{\phi\phi}}{\partial t}\right|^2\right\},
\end{eqnarray}
This expression is only valid in an asymptotically flat (AF) gauge,
by which we mean, following \cite{Cunningham:1979px}, 
\begin{eqnarray}\label{AFgauge1}
\pert{n}{h}_{tt}, \pert{n}{h}_{rr}, \pert{n}{h}_{tr} &=& O(r^{-2}),\\
\pert{n}{h}_{t\theta}, \pert{n}{h}_{t\phi}, \pert{n}{h}_{r\theta},
\pert{n}{h}_{r\phi} &=& O(r^{-1}),\\
\pert{n}{h}_{\theta\theta}, \pert{n}{h}_{\phi\phi},
\pert{n}{h}_{\theta\phi} &=& O(r),\\\label{AFgauge3}
\gamma^{ab}\pert{n}{h}_{ab} &=&O(r^0)
\end{eqnarray}
to the desired order $n$. 

Once the Regge-Wheeler and Zerilli functions are known up to a given
order, one can reconstruct the perturbed metric components in an AF gauge
(we will turn to this below). Once those are known, we could in principle
use Eq.~(\ref{Landau}) to compute the radiated energy. 

It turns out that the RW gauge is not asymptotically flat. Because of
that we will need to (explicitly or implicitly, this will be discussed
later) make a gauge transformation from the RW gauge to an asymptotically
flat one. Once in that gauge, we will apply Eq.~(\ref{Landau}) in order
to obtain the radiated power.

It is useful to rewrite the Landau and Lifshitz formula (\ref{Landau})
in covariant way with respect to the coordinates on $S^2$, as follows.
We first introduce the trace-free projector to the sphere (recall that
$\gamma_{ab}$ is the standard, unit metric on the sphere), 
\begin{equation}
P_{ab}{}^{cd}\equiv
\gamma_a{}^c\gamma_b{}^d-\frac{1}{2}\gamma_{ab}\gamma^{cd},
\end{equation}
such that for any rank-two tensor $A_{cd}$ we have that
$P_{ab}{}^{cd}A_{cd}$ is trace free. It allows us to introduce the
projected trace-free metric on the sphere,
\begin{equation}
i_{ab}\equiv P_{ab}{}^{cd}{\tilde g}_{cd},
\end{equation}
so that formula (\ref{Landau}) can be rewritten in the following way: 
\begin{equation}
\frac{d{\rm Power}}{d\Omega}
=\frac{1}{32\pi r^2} \left(\frac{\partial{i}_{ab}}{\partial t}\right)
\gamma^{ac}\gamma^{bd}
\left(\frac{\partial{i}_{cd}}{\partial t}\right)^*,
\end{equation}
where the star denotes complex conjugation.

Using the perturbative expansion in spherical harmonics
(\ref{metricdecomposition}), the projected trace-free metric takes
the form 
\begin{equation}
i_{ab}\!=\!\sum_{n=1}^{\infty}\!\frac{\varepsilon^n}{n!}\!\sum_{l,m}
\{
r^2\pert{n}{G}_l^m{}^{\rm AF}Z_l^m{}_{ab} + \!\pert{n}{h}_l^m{}^{\rm AF}X_l^m{}_{ab}
\},
\end{equation}
where the superscript AF stands for any asymptotically flat gauge.

Finally, making use of the normalization shown in appendix
\ref{tensorharmonics} and the fact that the tensor spherical harmonics
are trace-free, it is easy to integrate the emitted power over the solid
angle to get 
\begin{widetext}
\begin{eqnarray}\label{power}
{\rm Power} = \frac{1}{64\pi r^2}\sum_{j=1}^{\infty}\sum_{k=1}^{\infty}
\frac{\varepsilon^{j+k}}{j!k!}\sum_{l,m}
\frac{(l+2)!}{(l-2)!}
\left\{
r^4\left(\frac{\partial \pert{j}{G}_l^m{}^{\rm AF}}{\partial t}\right)
\left(\frac{\partial \pert{k}{G}_l^m{}^{\rm AF}}{\partial t}\right)^*
+\left(\frac{\partial \pert{j}{h}_l^m{}^{\rm AF}}{\partial t}\right)
\left(\frac{\partial \pert{k}{h}_l^m{}^{\rm AF}}{\partial t}\right)^*
\right\}.
\end{eqnarray}
\end{widetext}

Therefore, the problem of extracting the radiated power with an order
of $\varepsilon^{n}$ reduces to finding the value of the time derivative
of the harmonic coefficients $\pert{k}{G}_l^m$ and
$\pert{k}{h}_l^m$, for all $k < n$, in an asymptotically flat gauge. 
Note that in the last formula there is no coupling between modes with
different harmonic labels. This is so because of the integrated
character of the total emitted power, and the orthogonality between
different spherical harmonics. This has important consequences when one
wants to obtain the radiated power up to a given order $\varepsilon^n$
consistently, as discussed later.

\section{First order perturbations} \label{sec:first}

We now turn our attention to the reconstruction of the metric components
from the {\em first-order} master functions, and the computation of the
radiated energy in terms of them. The reason for re-deriving these
results here is twofold. First, to fix the conventions that we will
use in the second order case, in which the first order perturbations
will appear as `source terms'. Second, to sketch in a less complicated
case the type of calculations presented in the next section for second
order perturbations. 
 
After writing down the first order master equations we reconstruct the
metric in the RW gauge. One can explicitly transform the result from
the latter to an arbitrary gauge. In fact, in order to use the Landau
and Lifshitz formula (\ref{Landau}) we must use an asymptotically flat
gauge. As we will see, the RW gauge is not asymptotically flat. Because
of this, following \cite{Gleiser:1998rw}, we will first perform
an explicit asymptotic gauge transformation from the RW gauge to an AF
one and afterwards apply the LL formula. 

In addition, following \cite{Garat:1999vr}, we use an alternative way
of computing the radiated energy, which exploits the gauge invariant
form of the master functions presented in previous section, instead of 
making an explicit asymptotic gauge transformation. 

In this section we will remove all harmonic labels, as well as the
$n=1$ labels, since all objects will be of first order and correspond
to a generic harmonic pair $(l,m)$.

\subsection{Even parity/polar sector}

At first order, the Zerilli equation is a wave equation without sources,
\begin{equation}\label{Zer1}
{\Psi}^{|A}{}_A-V_{\rm Z} {\Psi}=0.
\end{equation}
This master variable contains the physical information of the system since
it is possible to reconstruct from it all components of the perturbation
of the metric, in the RW gauge. We explicitly display the results of such
reconstruction:
\begin{eqnarray}\label{FromZerilli1}
H_{tt}&=& \frac{2 M-r}{4 l (l+1) r^3 \left(3 M+\lambda r\right)^2}\times
\\\nonumber
&\left\{ \right.&\left.
2 (2 M-r) r^2 \left(6 M+ 2\lambda r\right)^2 \Psi''
\right.\\\nonumber
&+&\left. 4 r \left[\lambda
\left(l^2+l-8\right) r^2M
-2\lambda^2 r^3-18 M^3\right] \Psi'
\right.\\\nonumber
&+&\left.4\!\left[18M^3\!+\!18\lambda r M^2\!+\!
6\lambda^2 r^2 M\!+\! l (l+1)
\lambda^2 r^3\right]
\Psi\right\},\\
H_{rr}&=& \frac{r^2}{(2M-r)^2} H_{tt} ,\\
H_{tr}&=&\frac{2 \left(3 M^2+3 \lambda r M-\lambda r^2\right)}
{l (l+1) \left[6 M^2+\left(l^2+l-5\right) r M-\lambda r^2\right]} \dot{\Psi}
\nonumber\\
&+&\frac{2r}{l (l+1)} \dot{\Psi}',
\\\label{FromZerilli2}
K &=& \frac{1}{2 l(l+1) r^2 \left(3 M+\lambda r\right)}\times
\\\nonumber
&\left\{ \right.&\left.
2 r \left[-12 M^2-2\left(l^2+l-5\right) r M+2\lambda r^2\right] \Psi'
\right.\\\nonumber
&+&\left.\left[24 M^2+12 \lambda r M+ (l-1)l(l+1)(l+2) r^2\right] \Psi
\right\}.
\end{eqnarray}
Introducing these relations into the linearized Einstein equations,
one can show that all of them are trivially satisfied if the Zerilli
equation (\ref{Zer1}) holds.

Next we explicitly display the divergent nature of these quantities
(and, as a consequence, of the first order metric perturbations in the
RW gauge). For that purpose we temporarily introduce Schwarzschild
coordinates $(t,r)$ and the tortoise one $r^*$. The Zerilli function
$\Psi$ can be expanded in inverse powers of $r$, with coefficients
depending on the retarded time $u\equiv t-r^*$,
\begin{equation}\label{asintoticexpansion}
\Psi \equiv \Psi_0(u) + \frac{\Psi_1(u)}{r} + \frac{\Psi_2(u)}{r^2}
+ {\cal O}\left(\frac{1}{r^3}\right).
\end{equation}
The Zerilli equation is then equivalent to a series of relations, order
by order in $r$, among those coefficients:
\begin{eqnarray}
\Psi_0(u) = \frac{2}{l(l+1)} \ddot{F}(u),\qquad \Psi_1(u)
= \dot{F}(u),\nonumber\\
\Psi_2(u) = \frac{\lambda}{2} F(u) 
-\frac{3 M (\lambda +2)}{2\lambda(\lambda + 1)} \dot{F}(u), 
\end{eqnarray}
where $\dot{F}(u) = dF(u)/du$. The function $F(u)$ can be understood as
the free data at null infinity.

In order to see the divergent behaviour of the harmonic coefficients
in RW gauge at null infinity, we replace the expansion
(\ref{asintoticexpansion}) in (\ref{FromZerilli1}-\ref{FromZerilli2})
to obtain
\begin{eqnarray}
H_{tt}&=& \frac{4 \ddddot{F}}{l^2(l+1)^2} r
+\frac{4\lambda \dddot{F}} {l^2 (l+1)^2} +  {\cal O}\left(\frac{1}{r}\right),\\
H_{rr}&=& \frac{4 \ddddot{F}}{l^2(l+1)^2} r
+\frac{16M \ddddot{F}+4\lambda \dddot{F}}{l^2 (l+1)^2} +  {\cal O}\left(\frac{1}{r}\right) ,\\
H_{tr}&=& -\frac{4 \ddddot{F}}{l^2(l+1)^2} r
-\frac{8M \ddddot{F}+4\lambda\dddot F}{l^2 (l+1)^2} +  {\cal O}\left(\frac{1}{r}\right),\\
K &=& -\frac{4 {\dddot{F}}}{l^2(l+1)^2} r + {\cal O}\left(\frac{1}{r^2}\right),
\end{eqnarray}
where the orders $r^0$ and $r^{-1}$ vanish for the harmonic coefficient
$K$.

In order to apply the LL formula we perform an explicit asymptotic
(that is, near null infinity) transformation from the RW gauge to an AF
one. We will not show the form of the resulting change of coordinates but
instead directly show the asymptotic form of the metric coefficients in
the new gauge,
\begin{eqnarray}\label{asymptoticallyflat1}
H_{tt}^{\rm{AF}} &=& 0 + {\cal O}\left(\frac{1}{r^3}\right),\\
H_{rr}^{\rm{AF}}&=& 0 + {\cal O}\left(\frac{1}{r^3}\right),\\
H_{tr}^{\rm{AF}}&=& 0 + {\cal O}\left(\frac{1}{r^3}\right),\\
H_{t}^{\rm{AF}}&=& \left\{\dot\Phi_1
-\frac{1}{4l^2(l+1)^2}\left[4M\ddot{F}+\frac{(l+2)!}{(l-2)!}\dot{F}\right]
\right\}\frac{1}{r}
\nonumber\\
&+&{\cal O}\left(\frac{1}{r^2}\right),\\
H_{r}^{\rm{AF}}&=& -\left\{\dot\Phi_1
-\frac{1}{2l^2(l+1)^2}\left[2M\ddot{F}
\right.\right.\nonumber\\
&+&\left.\left.\lambda(l^2+l-8)\dot{F}\right]
\right\}\frac{1}{r}+{\cal O}\left(\frac{1}{r^2}\right),\\\label{Gasintotic}
G^{\rm{AF}}&=& \frac{4\ddot{F}}{l^2 (l+1)^2}\frac{1}{r}
+\frac{4\lambda \dot{F}}{l^2 (l+1)^2}\frac{1}{r^2}
+2{\Phi_1}\frac{1}{r^3}
\nonumber\\
&+&{\cal O}\left(\frac{1}{r^4}\right),\\\label{asymptoticallyflat7}
K^{\rm{AF}}&=& 0 + {\cal O}\left(\frac{1}{r^3}\right),
\end{eqnarray}
where zeros stand to show that in fact one could ask for faster
decay rates than the ones defined in
(\ref{AFgauge1}-\ref{AFgauge3})  and $\Phi_1=\Phi_1(u)$ is a
gauge freedom that it is not fixed by the requirement of
asymptotic flatness. From the behaviour of the harmonic
coefficient $G$ in an asymptotically flat gauge (\ref{Gasintotic})
and the asymptotic expansion of the Zerilli function
(\ref{asintoticexpansion}), it is easy to obtain
\begin{equation}\label{GPsi}
G^{\rm AF}=\frac{2\Psi}{l(l+1)r}+{\cal O}\left(\frac{1}{r^2}\right).
\end{equation}

Alternatively, this last relation can be directly obtained from the gauge
invariant definition of the Zerilli variable
(\ref{def1}-\ref{def2}). Since that definition is valid for any
gauge we can suppose that we are in an AF gauge. Imposing the
decay rates (\ref{AFgauge1}-\ref{AFgauge3}) it is straightforward
to obtain (\ref{GPsi}). The advantage of this last method is
that we do not have to do an explicit asymptotic gauge transformation.
The disadvantage is, however, that we need to assume that
(\ref{AFgauge1}-\ref{AFgauge3}) is indeed possible.

Either way, using the relation (\ref{GPsi}) and the LL formula we obtain
the radiated power in terms of the Zerilli function,
\begin{equation}
{\rm Power} =
\frac{\varepsilon^2}{16\pi}\sum_{l,m}\frac{(l-1)(l+2)}{l(l+1)}
\left|\frac{\partial \Psi_l^m}{\partial t}\right|^2.
\end{equation}
Since this expression holds asymptotically, we can now forget that we have
used intermediate Schwarzschild-type coordinates to derive it, since it
will hold for any time coordinate which agrees with it at infinity.

\subsection{Odd parity/axial sector} \label{sec:odd1}

We proceed as in the even parity/polar sector. We first reconstruct the
metric from the RW scalar $\Pi$ satisfying the RW equation 
\begin{equation}
-\left[\frac{1}{2r^2}(r^4{\Pi})^{|A}\right]_{|A}+\frac{(l-1)(l+2)}{2}{\Pi}=0,
\end{equation}
in the RW gauge:
\begin{eqnarray}\label{frompi}
h_t &=& \frac{r^2}{2\lambda}(2M-r) (4\Pi+r\Pi') ,\\
h_r &=& \frac{r^5}{2\lambda(2M-r)} \dot{\Pi}.
\end{eqnarray}

Next we expand the master scalar in inverse powers of $r$ near
the asymptotic null infinity $(r\rightarrow\infty, u=const.)$. Since
$\tilde\Pi=r^3\Pi$ obeys a standard wave equation,
our master scalar will have the following behavior,
\begin{equation}
\Pi\equiv \frac{\Pi_0(u)}{r^3} + \frac{\Pi_1(u)}{r^4} + \frac{\Pi_2(u)}{r^5} + {\cal O}\left(\frac{1}{r^6}\right),
\end{equation}
We can define a function $J(u)$ such that,
\begin{eqnarray}
\Pi_0(u) = \frac{2}{l(l+1)}\ddot J(u), \qquad \Pi_1(u) = \dot J(u),
\nonumber\\
\Pi_2(u) = \frac{\lambda}{2} J(u) -\frac{3 M}{l(l+1)}\dot J(u).
\end{eqnarray}

With these expansions at hand, we can obtain the precise divergent
behaviour of the RW gauge in terms of the function $J(u)$,
\begin{eqnarray}
h_t &=& \frac{r}{\lambda l(l+1)} \dddot{J}
+\frac{1}{l(l+1)}\ddot{J} + {\cal O}\left(\frac{1}{r}\right),\\
h_r &=& \!\!-\frac{r}{\lambda l(l+1)} \!\dddot{J}
-\!\frac{2 M}{\lambda l (l+1)} \!\dddot{J}
-\!\frac{1}{2\lambda}\ddot{J} + \!{\cal O}\left(\frac{1}{r}\right)\!.
\end{eqnarray}

As in the even parity/polar sector, we make an explicit asymptotic gauge
transformation to an AF gauge. And, as in that sector, we do not show the
details of the resulting transformation but instead the final asymptotic
behavior of the metric in the new gauge:
\begin{eqnarray}\label{axialasintotic1}
h_t^{AF} &=& \left\{ \dot{\Xi}_0
+\frac{1}{4}\dot{J}+\frac{(l-2)!}{(l+2)!}M\ddot{J}\right\}\frac{1}{r}
+ {\cal O}\left(\frac{1}{r^2}\right),\\
h_r^{AF} &=& -\left\{\dot{\Xi}_0
+\frac{(l-2)!}{(l+2)!}\left[
M\ddot{J}+\frac{\lambda}{2}(l^2+l-8)\dot{J}
\right]\right\}\frac{1}{r}\nonumber\\
&+& {\cal O}\left(\frac{1}{r^2}\right),\\\label{axialasintotic3}
h^{AF} &=&\!\!-\frac{2r}{\lambda l (l+1)} \ddot{J}- \frac{2}{l(l+1)} \dot{J}
+\frac{2}{r}\Xi_0+ \!{\cal O}\left(\frac{1}{r^2}\right)\!,
\end{eqnarray}
where $\Xi_0=\Xi_0(u)$ is a residual gauge freedom.
From there it is easy to obtain that asymptotically,
\begin{equation}\label{hPi}
h^{AF} = -\frac{r^4}{\lambda} \Pi + {\cal O}(r^0).
\end{equation}
Replacing this result in the LL formula for the emitted power we obtain
\begin{equation}
{\rm Power} = \frac{\varepsilon^2 r^6}{16\pi}\sum_{l,m}
\frac{l(l+1)}{(l-1)(l+2)}
\left|\frac{\partial \Pi_l^m}{\partial t}\right|^2.
\end{equation}

One can try to obtain relation (\ref{hPi}) with a gauge invariant approach,
as we have done in the polar case. It is not possible in this sector,
however, since the gauge invariant form of the master variable
$\Pi$ (\ref{defPi}) does not contain the harmonic coefficient $h$. Hence,
if using that variable one has to go through the explicit gauge
transformation. But at this point we note that there is another master
variable $\Phi$ whose gauge-invariant form (\ref{defPhi}) does contain
the harmonic coefficient $h$. Making a transformation to outgoing
coordinates and assuming that we are in an AF gauge we can easily obtain
between $\Phi$ and $h$ at null infinity,
\begin{equation}
\dot{h}^{\rm AF} =2 r \Phi + {\cal O}(r^0).
\end{equation}
Therefore, the emitted power can also be given in terms of this
last variable,
\begin{equation}
{\rm Power} = \frac{\varepsilon^2}{16\pi}\sum_{l,m}
\frac{(l+2)!}{(l-2)!}\left|\Phi_l^m\right|^2.
\end{equation}

Because we can apply this gauge-invariant approach to relate the
master variable with the harmonic coefficient $h$ at null infinity,
at second-order we will use the variable $\pert{2}{\Phi}$. But
there is one disadvantage of using $\Phi$ instead of $\Pi$. We
have shown above the reconstruction of the perturbations of the
metric in the RW gauge in terms of $\Pi$ (\ref{frompi}). These
relations are algebraic. If we try to do the same with the
variable $\Phi$, we find that the reconstruction of the metric
is not algebraic, but differential.
\begin{eqnarray}
h_r &=& \frac{r^2}{(r-2 M)}\Phi,\\
\dot{h_t} &=& \left(1-\frac{2 M}{r}\right)^2
   \left(h_r'+\frac{2 M}{r-2M}\frac{h_r}{r}\right).
\end{eqnarray}
This is why we will use $\Pi$ at linear order and $\Phi$ at second.

\section{Second order perturbations} \label{sec:second}

In order to solve for the second-order perturbations it is
{\em in principle} enough to solve the Zerilli (\ref{Zerilli}) and RW
(\ref{RW2eq}) equations with their corresponding sources [given by
(\ref{S2zerilli}) and (\ref{S2RW}), respectively]. However, as we will
discuss in the next two subsections, {\em in practice} there are some
technical obstacles to overcome first. 

\subsection{Sources: even parity sector}

As we have defined it, the second-order Zerilli function $\pert{2}{\Psi}$,
and in consequence also the source of the equation it obeys, 
$\pert{2}{S}_\Psi$, diverges at large radii. In order to see this, it is
sufficient to take its gauge-invariant definition (\ref{def1}-\ref{def2}),
assume an asymptotically flat gauge, and impose conditions
(\ref{asymptoticallyflat1}-\ref{asymptoticallyflat7}) and
(\ref{axialasintotic1}-\ref{axialasintotic3}). In this way, we find
that the quadratic source $\pert{2}{Q}_\Psi$ diverges as,
\begin{equation}
\pert{2}{Q}_\Psi = Q_2 r^2 + Q_1 r + Q_0 + {\cal O}\left(\frac{1}{r}\right),
\end{equation}
where $Q_0$, $Q_1$ and $Q_2$ are quadratic functions of
$\{\hat F,\bar F, \hat J, \bar J\}$. The hat and bar on $F$ and
$J$ denote the generic harmonic labels $(\hat l, \hat m)$
and $(\bar l, \bar m)$ of two first-order modes, respectively.
For instance, the dominant term is given by
\begin{equation}
Q_2 = \sum_{\hat l,\hat m}\sum_{\bar l,\bar m}
\frac{32}{\lambda \bar l^2(\bar l+1)^2 \hat l^2(\hat l+1)^2}
\E{0}{\hat l}{\hat m}{0}{\bar l}{\bar m}{l}
\ddddot{\hat F}\;\dddot{\overline F},
\end{equation}
where the $E$-coefficients, defined in Appendix \ref{tensorharmonics},
are normalized products of two Clebsch-Gordan factors.
In this case, the $E$ coefficient restricts the sums
to those harmonic labels $\{\hat l, \overline l, l\}$, such that
$\hat l + \overline l + l$ is an even number. That is,
for the cases in which $\hat l + \overline l + l$ is an odd
number the term $Q_2$ cancels out. In contrast, the term
$Q_1$ has a non-vanishing contribution in all cases.

These divergences are non-physical, and have their origin in the
freedom to add low-order gauge-invariant terms in the definition
of the second and higher order Zerilli functions, as discussed
in Section \ref{sec:high}.
Fortunately, they can be removed by exploiting that freedom to regularize
the source of the Zerilli equation.
In that way $\pert{n}{\Psi}$ will obey the same equation but with a
different, non diverging source term. The following discussion is rather
technical, but necessary. The reader not interested in the details,
though, might skip it and refer to Eq.~(\ref{Zereqreg}) [and
(\ref{RWeqreg}) for the odd parity sector], keeping in mind that we have
made use of the freedom in defining the second order master functions in
such a way that their associated sources are non-divergent both at the
horizon and at infinity. 

Our aim is to obtain some quadratic terms on the first-order Zerilli
functions and GS master scalars
$Q_{reg}=Q_{reg}[\{\hat\Psi,\overline\Psi,\hat\Pi,\overline\Pi\}]$
which reproduce the asymptotic divergent behavior of the source $Q_\Psi$
near null infinity. That is, for $r>>M$ with $u=const$,
\begin{eqnarray}\label{Qreginf}
Q_{reg}[\{\hat\Psi,\overline\Psi,\hat\Pi,\overline\Pi\}] &=&
Q_2[\{\hat F,\bar F, \hat J, \bar J\}] r^2 +
\\\nonumber
&&\hspace{-2.3cm}+ Q_1[\{\hat F,\bar F, \hat J, \bar J\}] r
+ Q_0[\{\hat F,\bar F, \hat J, \bar J\}]
+ {\cal O}\left(\frac{1}{r}\right).
\end{eqnarray}
In order to construct the function $Q_{reg}$ we make the following
replacements in $Q_2$, $Q_1$ and $Q_0$,
\begin{equation}\label{reconstruction}
\ddot{F}\to \frac{1}{2} l (l+1)\Psi,
\qquad\ddot{J}\to\frac{r^3}{2}l(l+1)\Pi.
\end{equation}
These rules include all cases but the first and zeroth
derivatives. There are no $F$ or $J$ terms without derivatives
in the divergent terms, but there are some first-order derivatives.
Hence, the straightforward definition would be
\begin{equation}
\dot{F}\to-r^2\left(\frac{\partial\Psi}{\partial r}\right)_u. \label{eq:straight}
\end{equation}
However, these replacements introduce divergences at the horizon $r=2M$.
In order to see this, we choose ingoing Eddington-Finkelstein
coordinates, which are smooth at the horizon. They are
obtained from the Schwarzschild coordinates $(t,r)$
by the following transformation
\begin{equation}
t\rightarrow w\equiv t + 2M \ln\left|\frac{r}{2M}-1\right|.
\end{equation}
In these coordinates the two-dimensional background metric takes the form 
\begin{eqnarray}
g_{AB}dx^Adx^B &=& -\left(1-\frac{2M}{r}\right)dw^2 +\frac{4M}{r}dwdr
\nonumber\\
&+&\left(1+\frac{2M}{r}\right)dr^2.
\end{eqnarray}
Therefore, we have the following relation between coordinate
derivatives,
\begin{equation}\label{totaylorexpand}
\left(\frac{\partial\Psi}{\partial r}\right)_u=
\left(\frac{\partial\Psi}{\partial r}\right)_\omega
+\frac{r+2 M}{r-2M}\left(\frac{\partial\Psi}{\partial\omega}\right)_r,
\end{equation}
which makes explicit the divergence of the radial derivative in
outgoing coordinates at the horizon $r=2 M$. Taking into account
Eq. (\ref{eq:straight}) this implies that the source diverges there
as well. 

In order to regularize the source at large radii without introducing
divergences at the horizon we proceed in the following way. 
First, we make a Taylor expansion in inverse powers of $r$ of the
right-hand side of Eq.~(\ref{totaylorexpand}). Next, we define a
derivative that approaches $(\partial/\partial r)_u$ for large $r$,
but without being divergent at the horizon. Following this method we get 
\begin{eqnarray}
\dot{F}\!=\!\!
-r^2\left(\frac{\partial\Psi}{\partial r}\right)_\omega
\!\!\!\!-\!(r^2\!\!+4Mr+8M^2)\!\left(\frac{\partial\Psi}{\partial\omega}\right)_r
\!\!\!\!+\!{\cal O}\!\left(\frac{1}{r}\right)\!\!,
\end{eqnarray}
which is finite at the horizon. Converting this last relation into
Schwarzschild coordinates gives the following rules to reconstruct
the divergent terms,
\begin{eqnarray}\label{reconstruction2}
\dot{F}\to-r^2\Psi'+\frac{r^3-16 M^3}{2 M-r}\dot{\Psi},\\
\dot{J}\to -r^2\left(r^3\Pi\right)'+\frac{r^3-16 M^3}{2 M-r}r^3\dot{\Pi}.
\end{eqnarray}

The replacements (\ref{reconstruction}) and (\ref{reconstruction2}) must
be done systematically. That is, first take $Q_2r^2$ and reconstruct
the term that will reproduce it,
\begin{equation}
\sum_{\hat l,\hat m}\sum_{\bar l,\bar m}
\frac{8 r^2}{\lambda \bar l(\bar l+1) \hat l(\hat l+1)}
\E{0}{\hat l}{\hat m}{0}{\bar l}{\bar m}{l}
\ddot{\hat \Psi}\dot{\overline \Psi}.
\end{equation}
When expanding near null infinity, this term will go
as $Q_2 r^2 + R_1 r + R_0 + {\cal O}(r^{-1})$. In order to remove
the divergent terms of order ${\cal O}(r)$, it is not enough to
find a term that will reproduce $Q_1r$, it must reproduce
$(Q_1-R_1)r$, to compensate the new term we have just introduced.
Therefore, we take $(Q_1-R_1)r$ and make the above
replacements (\ref{reconstruction}) and (\ref{reconstruction2}) again.
And so on, until we achieve the desired quadratic function
$Q_{reg}[\{\hat\Psi,\overline\Psi,\hat\Pi,\overline\Pi\}]$
which asymptotically behaves as in Eq.~ (\ref{Qreginf}).

In this way, we define the regularized second-order Zerilli function as
\begin{equation}
\pert{2}{\Psi}_{reg} \equiv \pert{2}{\Psi} +Q_{reg}.
\end{equation}
It obeys the following wave equation,
\begin{equation}\label{Zereqreg}
\pert{2}{\Psi_{reg}{}_l^m}_{|A}{}^{|A} - V_{\rm Z} \pert{2}{\Psi_{reg}{}_l^m} = \pert{2}{\cal S}^{reg}_\Psi,
\end{equation}
where the regularized source is given by
\begin{equation}
\pert{2}{\cal S}^{reg}_\Psi \equiv \pert{2}{\cal S}_\Psi+Q_{reg}{}_{|A}{}^{|A}-V_{\rm Z}Q_{reg}.
\end{equation}

We have implemented this regularization procedure for generic $(l\geq 2)$
first and second order modes, but the results are quite lengthy. Just to
illustrate the point, though, we explicitly show the final result for
the regularization factor for the particular case
$(\hat l, \hat m)=(\bar l, \bar m)=(l, m)=(2,0)$:
\begin{eqnarray}
Q_{reg}&=& - \frac{1}{252 \left(2M-r\right)}\sqrt{\frac{5}{\pi}}\times
\\\nonumber
& \left\{ \right.&\left.
2(2 M-r)\left((9 M+r)\pert{1}{\dot{\Psi}}+6\pert{1}{\Psi}\right) \pert{1}{\dot{\Psi}}
\right.\\\nonumber
&+&\left. \left(110 M^3-21 r M^2
+14 r^2 M+4 r^3\right)\pert{1}{\dot{\Psi}}\pert{1}{\ddot{\Psi}}
\right.\\\nonumber
&-&\left. 2 (2 M-r)\left(4r^2\pert{1}{\Psi}'
-(15 M-6 r) \pert{1}{\Psi}\right)\pert{1}{\ddot{\Psi}}\right\}\\\nonumber
&-&\frac{3r^6}{224}\sqrt{\frac{5}{\pi}}\left\{16\pert{1}{\dot{\Pi}}\pert{1}{\Pi}
+(2r-3M)\pert{1}{\dot{\Pi}}\pert{1}{\dot{\Pi}}\right\} .
\end{eqnarray}

\subsection{Sources: odd parity sector}

In the axial case there is no such a divergence. Following the same
steps as above, one finds that near null infinity the quadratic part
of the RW function $\pert{2}{\Phi}$ tends to
\begin{equation}
\pert{2}{Q}_\Phi={Q}_\Phi^{(0)} + {\cal O}\left(\frac{1}{r}\right).
\end{equation}
Therefore, in principle there is no need to regularize the second-order
RW function. But, as it will be clear in the next subsection, we are still
interested in removing the term of order ${\cal O}(1)$. We do so by
applying the same procedure as in the polar case: namely, we obtain a
term $Q_\Phi^{reg}$ which reproduces ${Q}_\Phi^{(0)}$ at null infinity. 

After that we define the regularized second-order RW variable as
\begin{equation}
\pert{2}{\Phi}_{reg} \equiv \pert{2}{\Phi} +Q^{reg}_\Phi,
\end{equation}
and its evolution equation
\begin{equation}\label{RWeqreg}
\pert{2}{\Phi_{reg}{}_l^m}_{|A}{}^{|A} - V_{\rm RW} \pert{2}{\Phi_{reg}{}_l^m} = \pert{2}{\cal S}^{reg}_\Phi,
\end{equation}
where the regularized source is again given by
\begin{equation}
\pert{2}{\cal S}^{reg}_\Phi \equiv \pert{2}{\cal S}_\Phi+Q^{reg}_\Phi{}_{|A}{}^{|A}-V_{\rm RW}Q^{reg}_\Phi.
\end{equation}

As in the even parity case, we shall not present the details of the
general procedure but instead simply explicitly show the final result
for the regularization factor for the
$(\hat l, \hat m)=(\bar l, \bar m)=(l, m)=(2,0)$ case:
\begin{equation}
Q_\Phi^{reg} = -\frac{r^3}{84}\sqrt{\frac{5}{\pi}}\{3\dot{\Pi}\dot{\Psi}
+\ddot{\Pi}\Psi+\ddot{\Psi}\Pi\}.
\end{equation}

The regularized sources for the equations of motion (\ref{Zereqreg}) and
(\ref{RWeqreg}) are one of the main results of this article. We have
calculated them for the presence of any first- and second-order axial or
polar modes. We do not include their explicit form here because they are
quite lengthy and do not contribute to the discussion. They are available
from the authors upon request. As an illustration, for the interested
reader, in Appendix \ref{sources} we have written down their explicit
form for some particular values of the harmonic labels.

\subsection{Radiated power}

Once we have solved for the first and second-order master equations we
can obtain the radiated power by using Eq.~(\ref{power}). Expanding it
explicitly up to order $\varepsilon^3$, it takes the following form
\begin{widetext}
\begin{eqnarray}
{\rm Power} &=& \frac{\varepsilon^2}{64\pi r^2}\sum_{l,m}\frac{(l+2)!}{(l-2)!}
\left\{
r^4\left|\frac{\partial \pert{1}{G}_l^m{}^{AF}}{\partial t}\right|^2
+\left|\frac{\partial \pert{1}{h}_l^m{}^{AF}}{\partial t}\right|^2
+\varepsilon Re\left[r^4\frac{\partial \pert{1}{G}_l^m{}^{AF}}{\partial t}
\left(\frac{\partial \pert{2}{G}_l^m{}^{AF}}{\partial t}\right)^*
\right.\right.\nonumber\\
&+& \left.\left.\frac{\partial \pert{1}{h}_l^m{}^{AF}}{\partial t}
\left(\frac{\partial \pert{2}{h}_l^m{}^{AF}}{\partial t}\right)^*\right]
\right\}+{\cal O}(\varepsilon^4),\label{eq:power3}
\end{eqnarray}
\end{widetext}
where $Re$ means the real part. Again, the problem of finding the
radiated power reduces to calculating the harmonic coefficients $G_l^m$
and $h_l^m$ near null infinity, in an asymptotically flat gauge.
More precisely, we want to relate them with the regularized master
scalars constructed in the previous two subsections. 

In those subsections we regularized the second-order master variables so
that the quadratic contributions from first-order modes decay as
${\cal O}(1/r)$ near null infinity. Hence, we can use their gauge-invariant
definitions, (\ref{def2}) and (\ref{defPi}), and assume an AF gauge
(\ref{AFgauge1}-\ref{AFgauge3}) up to second order.
This leads to the very same relations as at first-order; namely,
\begin{eqnarray}
\pert{2}{G_l^m}{}^{\rm AF} &=& \frac{2\pert{2}{\Psi}_l^m{}_{reg}}{l(l+1)r} + {\cal O}\left(\frac{1}{r^2}\right),\\
\pert{2}{\dot{h}_l^m}{}^{\rm AF} &=&2 r \pert{2}{\Phi}_l^m{}_{reg} + {\cal O}(r^0).
\end{eqnarray}
Replacing these expressions in Eq.~(\ref{eq:power3}), the radiated power 
up to order $\varepsilon^3$ is given in terms of the master scalars by
\begin{widetext}
\begin{eqnarray}
{\rm Power} &=& \frac{\varepsilon^2}{64\pi}\sum_{l,m}\frac{(l+2)!}{(l-2)!}
\left\{
\frac{4}{l^2(l+1)^2}\left|\frac{\partial \pert{1}{\Psi}_l^m}{\partial t}\right|^2
+\frac{r^6}{\lambda^2}\left|\frac{\partial \pert{1}{\Pi}_l^m}{\partial t}\right|^2
+\varepsilon Re\left[\frac{4}{l^2(l+1)^2}\frac{\partial \pert{1}{\Psi}_l^m}{\partial t}
\left(\frac{\partial \pert{2}{\Psi}_l^m{}_{reg}}{\partial t}\right)^*
\right.\right.\nonumber\\
&-& \left.\left.\frac{2r^3}{\lambda}\frac{\partial \pert{1}{\Pi}_l^m}{\partial t}
\pert{2}{\Phi}_l^m{}_{reg}^*\right]
\right\}+{\cal O}(\varepsilon^4).
\end{eqnarray}
\end{widetext}
This last formula, complemented with the evolution equations for the
regularized master scalars constructed above, provides a closed set of
formulas that permits to obtain the radiated power up to order
$\varepsilon^3$ in the most general case in a fully consistent way.

At this point we want to discuss an aspect of second order perturbations
of Schwarzschild black holes which seems not have been discussed before
in the literature. 

Namely, even when we are solving for the metric up to second-order
perturbations, we can only obtain the {\em complete} radiated power up
to third-order in $\varepsilon$. In order to obtain the following order
$\varepsilon^4$, one should also consider third-order perturbations.

Let us further elaborate on this point. Consider the simplest possible
scenario: a unique first-order mode with harmonic labels $(l, m)$ and
polarity $\sigma$. The polarity $\sigma$ will take the value $1$ for
even-parity/polar modes and $-1$ for odd-parity/axial modes. Because
of reality conditions, if the mode $(l,m,\sigma)$ is present, so is
its conjugated $(l,-m,\sigma)$. 

The self-coupling of this mode will generate several second-order modes
but not in general the one with labels $(l,\pm m,\sigma)$.
In contrast, at third-order the mode with indices $(l,\pm m,\sigma)$ 
will indeed be generated. This means that the third-order modes will always
contribute to the emitted power at order $\varepsilon^4$, coupled
to the first-order mode with the same harmonic labels. Therefore,
without considering third-order modes, one can only obtain the
radiated power consistently up to order $\varepsilon^3$. 

In order for the emitted power (\ref{power}) to have a contribution of that
order $(\varepsilon^3)$ the self-coupling of the first-order mode
must give a second-order mode with the same labels $(l,m,\sigma)$.
It is easy to see that, when only considering a first-order mode
$(l,\pm m,\sigma)$, this will happen if and only if $m=0$ and if,
for $\sigma=1$ ($\sigma=-1$), $l$ is an even (odd) number. 

In order to make the above discussion more explicit and analyze which
problems can be addressed consistently, let us consider the
particular case of a first-order even-parity/polar mode with harmonic
labels $l=m=2$ and $l=m=-2$. These modes will generate the
second-order $\{l=4,m=\pm 4,0\}$, $\{l=2,m=0\}$ and $\{l=0,m=0\}$
even-parity/polar modes as well as the $\{l=3,m=0\}$ odd-parity/axial mode.
Particularizing the power formula (\ref{power}) in terms of
the master scalars to this case, we obtain the following
contributions from the modes,
\begin{widetext}
\begin{equation}
{\rm Power} =
\frac{\varepsilon^2}{12\pi}\left|\partial_t\pert{1}{\Psi}_2^2\right|^2
+\frac{9\varepsilon^4}{640\pi}\left\{\left|\partial_t\pert{2}{\Psi}_4^0\right|^2
+2\left|\partial_t\pert{2}{\Psi}_4^4\right|^2
\right\}
+\frac{15\varepsilon^4}{8\pi}\left|\pert{2}{\Phi}_3^0\right|^2
+\frac{\varepsilon^4}{96\pi}\left|\partial_t\pert{2}{\Psi}_2^0\right|^2,
\end{equation}
\end{widetext}
where the second-order master scalars are the regularized ones.
Here it can be clearly seen that the order $\varepsilon^3$ is not
present. The problem with this last formula is that it is not
complete since the third-order $\{l=2,m=\pm 2\}$ polar mode
would contribute to the power at order $\varepsilon^4$.

On the other hand, let us consider the first-order mode $l=2$ with
all its possible harmonic labels $m=0,\pm 1, \pm 2$. By coupling,
they will generate the second-order polar modes $l=0$, $l=2$ and
$l=4$ with all their possible $m$. That is, we will have the
second-order $\{l=0,m=0\}$, $\{l=2,m=0,\pm 1, \pm 2\}$ and
$\{l=4,m=0,\pm 1, \pm 2,\pm 3, \pm 4\}$ polar modes. This
particular case will provide a non-vanishing $\varepsilon^3$-order
term to the power,
\begin{widetext}
\begin{eqnarray}
{\rm Power} &=&
\frac{\varepsilon^2}{24\pi}\left\{2\left|\partial_t\pert{1}{\Psi}_2^2\right|^2
+2\left|\partial_t\pert{1}{\Psi}_2^1\right|^2
+\left|\partial_t\pert{1}{\Psi}_2^0\right|^2\right\}
\nonumber\\
&& +\frac{\varepsilon^3}{24\pi} Re\left[
2\partial_t(\pert{1}{\Psi}_2^2)\partial_t(\pert{2}{\Psi}_2^2)^*
\nonumber\right.
+\left.2\partial_t(\pert{1}{\Psi}_2^1)\partial_t(\pert{2}{\Psi}_2^1)^*
+\partial_t(\pert{1}{\Psi}_2^0)\partial_t(\pert{2}{\Psi}_2^0)^*
\right]
\nonumber\\
&&+{\cal O}(\varepsilon^4),
\end{eqnarray}
\end{widetext}
where, again, the second-order Zerilli function must be understood
as regularized. 
In this last case the formula is exact up to the displayed order,
to which the generated second-order axial modes and third-order
polar modes do not contribute.

\section{Final remarks} \label{sec:comments}

In this paper we have introduced a complete gauge invariant formalism to
study arbitrary perturbations of a Schwarzschild black hole up to second
order. In particular, we regularized the resulting equations, making them 
suitable for a numerical implementation. 

This formalism enables a variety of applications and studies. These range
from the non-linear stability of the black hole horizon, to non-linear
features in gravitational waves and mode-mode coupling. We will report
those studies elsewhere. 

All calculations of this paper have been done with --and have been largely
possible at all due to-- new, efficient symbolic manipulations tools.
The resulting expressions in most cases are very long and their
explicit expressions are not particularly enlightening. For that reason
we have refrained from explicitly presenting most of them. They are
however, available upon request.

\acknowledgments
We thank Vitor Cardoso, Jorge Pullin, and Olivier Sarbach for helpful comments on the manuscript. 

This research has been supported in part by 
the Spanish MICINN Project FIS2008-06078-C03-03,
the French A.N.R. Grant {\it LISA Science} BLAN07-1\_201699
and NSF Grant No. 0801213 to the University of Maryland.
DB has been supported by the FPI program of the Regional Government of
Madrid.

\bibliography{references}

\begin{thebibliography}{20}
\expandafter\ifx\csname natexlab\endcsname\relax\def\natexlab#1{#1}\fi
\expandafter\ifx\csname bibnamefont\endcsname\relax
  \def\bibnamefont#1{#1}\fi
\expandafter\ifx\csname bibfnamefont\endcsname\relax
  \def\bibfnamefont#1{#1}\fi
\expandafter\ifx\csname citenamefont\endcsname\relax
  \def\citenamefont#1{#1}\fi
\expandafter\ifx\csname url\endcsname\relax
  \def\url#1{\texttt{#1}}\fi
\expandafter\ifx\csname urlprefix\endcsname\relax\def\urlprefix{URL }\fi
\providecommand{\bibinfo}[2]{#2}
\providecommand{\eprint}[2][]{\url{#2}}

\bibitem[{\citenamefont{Brizuela et~al.}(2006)\citenamefont{Brizuela,
  Mart\'in-Garc\'ia, and Mena~Marug\'an}}]{Brizuela:2006ne}
\bibinfo{author}{\bibfnamefont{D.}~\bibnamefont{Brizuela}},
  \bibinfo{author}{\bibfnamefont{J.~M.} \bibnamefont{Mart\'in-Garc\'ia}},
  \bibnamefont{and} \bibinfo{author}{\bibfnamefont{G.~A.}
  \bibnamefont{Mena~Marug\'an}}, \bibinfo{journal}{Phys. Rev.}
  \textbf{\bibinfo{volume}{D74}}, \bibinfo{pages}{044039}
  (\bibinfo{year}{2006}), \eprint{gr-qc/0607025}.

\bibitem[{\citenamefont{Brizuela et~al.}(2007)\citenamefont{Brizuela,
  Mart\'{\i}n-Garc\'{\i}a, and Mena~Marug\'an}}]{Brizuela:2007zza}
\bibinfo{author}{\bibfnamefont{D.}~\bibnamefont{Brizuela}},
  \bibinfo{author}{\bibfnamefont{J.~M.} \bibnamefont{Mart\'{\i}n-Garc\'{\i}a}},
  \bibnamefont{and} \bibinfo{author}{\bibfnamefont{G.~A.}
  \bibnamefont{Mena~Marug\'an}}, \bibinfo{journal}{Phys. Rev.}
  \textbf{\bibinfo{volume}{D76}}, \bibinfo{pages}{024004}
  (\bibinfo{year}{2007}), \eprint{gr-qc/0703069}.

\bibitem[{\citenamefont{Brizuela et~al.}(2009)\citenamefont{Brizuela,
  Mart\'{\i}n-Garc\'{\i}a, and Mena~Marug\'an}}]{Brizuela:2008ra}
\bibinfo{author}{\bibfnamefont{D.}~\bibnamefont{Brizuela}},
  \bibinfo{author}{\bibfnamefont{J.~M.} \bibnamefont{Mart\'{\i}n-Garc\'{\i}a}},
  \bibnamefont{and} \bibinfo{author}{\bibfnamefont{G.~A.}
  \bibnamefont{Mena~Marug\'an}}, \bibinfo{journal}{Gen. Rel. Grav.}
  (\bibinfo{year}{2009}), \eprint{0807.0824}.

\bibitem[{\citenamefont{Mart\'{\i}n-Garc\'{\i}a}(2008)}]{xAct}
\bibinfo{author}{\bibfnamefont{J.~M.} \bibnamefont{Mart\'{\i}n-Garc\'{\i}a}},
  \bibinfo{journal}{Comp. Phys. Commun.} \textbf{\bibinfo{volume}{179}},
  \bibinfo{pages}{597} (\bibinfo{year}{2008}),
  \urlprefix\url{http://metric.iem.csic.es/Martin-Garcia/xAct/}.

\bibitem[{\citenamefont{Regge and Wheeler}(1957)}]{Regge57}
\bibinfo{author}{\bibfnamefont{T.}~\bibnamefont{Regge}} \bibnamefont{and}
  \bibinfo{author}{\bibfnamefont{J.}~\bibnamefont{Wheeler}},
  \bibinfo{journal}{Phys. Rev.} \textbf{\bibinfo{volume}{108}},
  \bibinfo{pages}{1063} (\bibinfo{year}{1957}).

\bibitem[{\citenamefont{Zerilli}(1970)}]{Zerilli70}
\bibinfo{author}{\bibfnamefont{F.~J.} \bibnamefont{Zerilli}},
  \bibinfo{journal}{Phys. Rev. Lett.} \textbf{\bibinfo{volume}{24}},
  \bibinfo{pages}{737} (\bibinfo{year}{1970}).

\bibitem[{\citenamefont{Moncrief}(1974)}]{Moncrief:1974vm}
\bibinfo{author}{\bibfnamefont{V.}~\bibnamefont{Moncrief}},
  \bibinfo{journal}{Ann. Phys. (N.Y.)} \textbf{\bibinfo{volume}{88}},
  \bibinfo{pages}{323} (\bibinfo{year}{1974}).

\bibitem[{\citenamefont{Tomita}(1974)}]{Tomita74}
\bibinfo{author}{\bibfnamefont{K.}~\bibnamefont{Tomita}},
  \bibinfo{journal}{Prog. Theor. Phys.} \textbf{\bibinfo{volume}{52}},
  \bibinfo{pages}{1188} (\bibinfo{year}{1974}).

\bibitem[{\citenamefont{Tomita and Tajima}(1976)}]{Tomita76}
\bibinfo{author}{\bibfnamefont{K.}~\bibnamefont{Tomita}} \bibnamefont{and}
  \bibinfo{author}{\bibfnamefont{N.}~\bibnamefont{Tajima}},
  \bibinfo{journal}{Prog. Theor. Phys.} \textbf{\bibinfo{volume}{56}},
  \bibinfo{pages}{551} (\bibinfo{year}{1976}).

\bibitem[{\citenamefont{{Price} and {Pullin}}(1994)}]{1994PhRvL..72.3297P}
\bibinfo{author}{\bibfnamefont{R.~H.} \bibnamefont{{Price}}} \bibnamefont{and}
  \bibinfo{author}{\bibfnamefont{J.}~\bibnamefont{{Pullin}}},
  \bibinfo{journal}{Physical Review Letters} \textbf{\bibinfo{volume}{72}},
  \bibinfo{pages}{3297} (\bibinfo{year}{1994}), \eprint{gr-qc/9402039}.

\bibitem[{\citenamefont{Gleiser et~al.}(2000)\citenamefont{Gleiser, Nicasio,
  Price, and Pullin}}]{Gleiser:1998rw}
\bibinfo{author}{\bibfnamefont{R.~J.} \bibnamefont{Gleiser}},
  \bibinfo{author}{\bibfnamefont{C.~O.} \bibnamefont{Nicasio}},
  \bibinfo{author}{\bibfnamefont{R.~H.} \bibnamefont{Price}}, \bibnamefont{and}
  \bibinfo{author}{\bibfnamefont{J.}~\bibnamefont{Pullin}},
  \bibinfo{journal}{Phys. Rept.} \textbf{\bibinfo{volume}{325}},
  \bibinfo{pages}{41} (\bibinfo{year}{2000}), \eprint{gr-qc/9807077}.

\bibitem[{\citenamefont{Nicasio et~al.}(2000)\citenamefont{Nicasio, Gleiser,
  and Pullin}}]{Nicasio:2000ge}
\bibinfo{author}{\bibfnamefont{C.~O.} \bibnamefont{Nicasio}},
  \bibinfo{author}{\bibfnamefont{R.}~\bibnamefont{Gleiser}}, \bibnamefont{and}
  \bibinfo{author}{\bibfnamefont{J.}~\bibnamefont{Pullin}},
  \bibinfo{journal}{Gen. Rel. Grav.} \textbf{\bibinfo{volume}{32}},
  \bibinfo{pages}{2021} (\bibinfo{year}{2000}), \eprint{gr-qc/0001021}.

\bibitem[{\citenamefont{Lousto and Nakano}(2009)}]{Lousto:2009}
\bibinfo{author}{\bibfnamefont{C.~O.} \bibnamefont{Lousto}} \bibnamefont{and}
  \bibinfo{author}{\bibfnamefont{H.}~\bibnamefont{Nakano}},
  \bibinfo{journal}{Class. Quant. Grav.} \textbf{\bibinfo{volume}{26}},
  \bibinfo{pages}{015007} (\bibinfo{year}{2009}), \eprint{0804.3824}.

\bibitem[{\citenamefont{Gerlach and Sengupta}(1979)}]{Gerlach79}
\bibinfo{author}{\bibfnamefont{U.~H.} \bibnamefont{Gerlach}} \bibnamefont{and}
  \bibinfo{author}{\bibfnamefont{U.~K.} \bibnamefont{Sengupta}},
  \bibinfo{journal}{Phys. Rev. D.} \textbf{\bibinfo{volume}{19}},
  \bibinfo{pages}{2268} (\bibinfo{year}{1979}).

\bibitem[{\citenamefont{Gerlach and Sengupta}(1980)}]{Gerlach:1980tx}
\bibinfo{author}{\bibfnamefont{U.~H.} \bibnamefont{Gerlach}} \bibnamefont{and}
  \bibinfo{author}{\bibfnamefont{U.~K.} \bibnamefont{Sengupta}},
  \bibinfo{journal}{Phys. Rev.} \textbf{\bibinfo{volume}{D22}},
  \bibinfo{pages}{1300} (\bibinfo{year}{1980}).

\bibitem[{\citenamefont{Sarbach and Tiglio}(2001)}]{Sarbach:2001qq}
\bibinfo{author}{\bibfnamefont{O.}~\bibnamefont{Sarbach}} \bibnamefont{and}
  \bibinfo{author}{\bibfnamefont{M.}~\bibnamefont{Tiglio}},
  \bibinfo{journal}{Phys. Rev.} \textbf{\bibinfo{volume}{D64}},
  \bibinfo{pages}{084016} (\bibinfo{year}{2001}), \eprint{gr-qc/0104061}.

\bibitem[{\citenamefont{Martel and Poisson}(2005)}]{Martel:2005ir}
\bibinfo{author}{\bibfnamefont{K.}~\bibnamefont{Martel}} \bibnamefont{and}
  \bibinfo{author}{\bibfnamefont{E.}~\bibnamefont{Poisson}},
  \bibinfo{journal}{Phys. Rev.} \textbf{\bibinfo{volume}{D71}},
  \bibinfo{pages}{104003} (\bibinfo{year}{2005}), \eprint{gr-qc/0502028}.

\bibitem[{\citenamefont{Garat and Price}(2000)}]{Garat:1999vr}
\bibinfo{author}{\bibfnamefont{A.}~\bibnamefont{Garat}} \bibnamefont{and}
  \bibinfo{author}{\bibfnamefont{R.~H.} \bibnamefont{Price}},
  \bibinfo{journal}{Phys. Rev.} \textbf{\bibinfo{volume}{D61}},
  \bibinfo{pages}{044006} (\bibinfo{year}{2000}), \eprint{gr-qc/9909005}.

\bibitem[{\citenamefont{Landau and Lifshitz}(1975)}]{Landau75}
\bibinfo{author}{\bibfnamefont{L.~D.} \bibnamefont{Landau}} \bibnamefont{and}
  \bibinfo{author}{\bibfnamefont{E.~M.} \bibnamefont{Lifshitz}},
  \emph{\bibinfo{title}{The Classical Theory of Fields}}
  (\bibinfo{publisher}{Pergamon Press}, \bibinfo{address}{Oxford},
  \bibinfo{year}{1975}).

\bibitem[{\citenamefont{Cunningham et~al.}(1979)\citenamefont{Cunningham,
  Price, and Moncrief}}]{Cunningham:1979px}
\bibinfo{author}{\bibfnamefont{C.~T.} \bibnamefont{Cunningham}},
  \bibinfo{author}{\bibfnamefont{R.~H.} \bibnamefont{Price}}, \bibnamefont{and}
  \bibinfo{author}{\bibfnamefont{V.}~\bibnamefont{Moncrief}},
  \bibinfo{journal}{Aptrophys J.} \textbf{\bibinfo{volume}{230}},
  \bibinfo{pages}{870} (\bibinfo{year}{1979}).

\end{thebibliography}

\begin{appendix}

\section{Tensor spherical harmonics}\label{tensorharmonics}

Tensor fields of any rank $s$ on the sphere will be decomposed using
a basis of tensor spherical harmonics. Such basis can be constructed from
the symmetric trace-free (STF) tensors
\begin{eqnarray}
Z_l^m{}_{a_1...a_s}&\equiv &\left(Y_l^m{}_{:a_1...a_s}\right)^{\rm STF},\\
X_l^m{}_{a_1...a_s}&\equiv &\epsilon_{(a_1}{}^b Z_l^m{}_{ba_2...a_s)},
\end{eqnarray}
together with the metric $\gamma_{ab}$ and the antisymmetric tensor
$\epsilon_{ab}$ \cite{Brizuela:2006ne}. For the particular case $s=0$, those
objects must be read as $Z_l^m\equiv Y_l^m$ and $X_l^m\equiv 0$.
They are normalized in the following way,
\begin{eqnarray}\label{normalization}
\int d\Omega Z_l^m{}_{ab}\gamma^{ac}\gamma^{bd}\left(Z_{l'}^{m'}{}_{cd}\right)^*&=&
\frac{1}{2}\frac{(l+2)!}{(l-2)!}\delta_{ll'}\delta_{mm'},\\
\int d\Omega X_l^m{}_{ab}\gamma^{ac}\gamma^{bd}\left(X_{l'}^{m'}{}_{cd}\right)^*&=&
\frac{1}{2}\frac{(l+2)!}{(l-2)!}\delta_{ll'}\delta_{mm'},\\
\int d\Omega X_l^m{}_{ab}\gamma^{ac}\gamma^{bd}\left(Z_{l'}^{m'}{}_{cd}\right)^*&=& 0.
\end{eqnarray}

Going beyond linear perturbation theory, the nonlinear coupling between
two first-order modes results in products between two tensor spherical
harmonics $(l,m,s)$ and $(l',m',s')$.
Those products can be decomposed into a linear combination of harmonics
$(l'',m+m',s+s')$
with an explicit formula involving coefficients
\begin{equation}
\E{s'}{l'}{m'}{s}{l}{m}{l''} \equiv
\frac{k(l',|s'|)k(l,|s|)}{k(l'',|s+s'|)} \C{l'}{m'}{l}{m}{l''}{m'+m}
\C{l'}{s'}{l}{s}{l''}{s'+s},
\end{equation}
where $\C{l'}{m'}{l}{m}{l''}{m''}$ are the usual Clebsch-Gordan coefficients
and $k$ is a normalization factor defined by,
\begin{equation}
k(l,s) = \sqrt{\frac{(2l+1)(l+s)!}{ \,\,2^{s+2}\,\pi\,(l-s)!}}.
\end{equation}
These $E$-coefficients encode the geometric selection rules that determine
which pairs of modes do actually couple. See \cite{Brizuela:2006ne} for full details.

\section{Regularized sources, some explicit examples}\label{sources}

In this appendix we show two particular examples for the regularized
sources of the RW (\ref{RWeqreg}) and Zerilli equations (\ref{Zereqreg}).

Let us first assume that we have the first-order $\{l=2,m=1\}$
even-parity/polar and $\{l=8,m=-4\}$ odd-parity/axial modes. 
The regularized source generated by them for the equation of motion
of the, for example, second-order $\{l=7,m=3\}$ even-parity/polar mode
is given by 
\begin{widetext}
\begin{eqnarray}\nonumber
S^{reg}_\Zer &=&-\frac{i \sqrt{\frac{22}{51 \pi }} r}{945(U+9)^2 (2 U-1) (3 U+2)^2} \Big\{-60 \Pi_{,tr} \left(6 U^3+55 U^2+7 U-18\right)^2r^3\Zer_{,rr}
\\\nonumber
&+&60 \Pi_{,rr} \left(6 U^3+55U^2+7 U-18\right)^2 r^3\Zer_{,tr}
\\\nonumber
&-&20\Pi_{,tr} (U+9)^2 \left(18 U^4+51 U^3-58U^2-26 U+20\right) r^2 \Zer_{,r}
\\\nonumber
&-&5\Pi_{,t} (3 U+2)^2 \left(212 U^4+3364U^3+11603 U^2-13149 U+3240\right)r^2 \Zer_{,rr}
\\\nonumber
&+&20 \Pi_{,rr} (U+9)^2 \left(126 U^4+141U^3-82 U^2-50 U+20\right) r^2 \Zer_{,t}
\\\nonumber
&+&5\Pi_{,r} (3 U+2)^2 \left(308 U^4+4972U^3+17255 U^2-22221 U+6156\right)r^2 \Zer_{,tr}
\\\nonumber
&-&180 \Pi_{,tr} (U+9)^2 \left(6 U^4+9 U^3+2U^2+4 U-4\right) r \Zer
\\\nonumber
&+&\Pi_{,t}\left(360 U^6+2568 U^5+57529 U^4-14036U^3+375894 U^2+88254 U
-123444\right) r\Zer_{,r}
\\\nonumber
&+&\Pi_{,r} \left(14220U^6+253302 U^5+1234181 U^4+854111 U^3
-966354U^2-375354 U +220644\right)  r\Zer_{,t}
\\\nonumber
&+&15\Pi (3 U+2)^2 \left(92 U^4+1492 U^3 +8507U^2+14154 U-9396\right)
r\Zer_{,tr}
\\\nonumber
&-&\Pi_{,t} \left(6840 U^6+112128 U^5+422069U^4
+1424 U^3-213006 U^2+271944 U-48924\right)\Zer
\\
&+&\Pi \left(6210 U^6+116313
U^5+1283789 U^4+6929894 U^3+6649074 U^2
-316926 U-1359504\right) \Zer_{,t}\Big\},
\end{eqnarray}
\end{widetext}
where the symbol $U$ stands for the dimensionless mass $U\equiv M/r$. 

As a second example, we show the regularized
source for the RW equation (\ref{RWeqreg}) for the
particular case in which the first-order even-parity/polar modes $(\bar l=3,\bar m=0)$
and $(\hat l=4,\hat m=-1)$ generate a second-order odd-parity/axial mode
with labels $(l=4,m=-1)$:
\begin{widetext}
\begin{eqnarray}\nonumber
S^{reg}_\RW &=&
\frac{3 i}{8800 \sqrt{7 \pi } (U+3)^4 (2 U-1)^2 (3U+5)^4}
\Big\{
-10 r \left(3 U^2+14 U+15\right)^4 \hat\Zer_{,rrr}\bar\Zer_{,rr} (2 U-1)^5
\\\nonumber
&+&26 r \left(3 U^2+14U+15\right)^4 \hat\Zer_{,rr} \bar\Zer_{,rrr}(2U-1)^5
\\\nonumber
&-&\frac{(3 U+5)^2}{r^2} \left(3060 U^8+49401U^7+332356 U^6+1197973 U^5+2636572 U^4
\right.\\\nonumber
&+&\left.3760905U^3+2764530 U^2-467775 U-1518750\right)\hat\Zer_{,rr} \bar\Zer(2U-1)^3
\\\nonumber
&+&\frac{(U+3)^2}{r^2} \left(1620 U^8+28161
U^7+173844 U^6+197637 U^5-1511900 U^4
\right.\\\nonumber
&-&\left.5534775
U^3-7023550 U^2-3510375 U-573750\right)
   \hat\Zer \bar\Zer_{,rr} (2 U-1)^3
\\\nonumber
&+&10 r\left(3 U^2+14 U+15\right)^4 \hat\Zer_{,trr}\bar\Zer_{,tr} (2 U-1)^3
-26 r \left(3 U^2+14U+15\right)^4 \hat\Zer_{,tr} \bar\Zer_{,trr} (2U-1)^3
\\\nonumber
&-&\frac{16 (1-2 U)^2}{r^4} \left(1701U^{11}+35262 U^{10}+320166 U^9+1720086U^8
+6285736 U^7
\right.\\\nonumber
&+&\left.16821825 U^6+34748135U^5+56990175 U^4+68601150 U^3+42931125U^2
\right.\\\nonumber
&-&\left.5703750 U-15946875\right) \hat\Zer\bar\Zer
\\\nonumber
&-&\frac{2}{r^3} \left(6 U^2+7
U-5\right)^2 \left(1530 U^9+34221 U^8+303099
U^7+1485635 U^6+4592169 U^5
\right.\\\nonumber
&+&\left.9179205 U^4+10353033
U^3+3316365 U^2-2994975 U-1478250\right)
   \hat\Zer_{,r} \bar\Zer
\\\nonumber
&-&\frac{10}{r} (U+3)^2
   \left(6 U^2+7 U-5\right)^4 \left(U^3+9 U^2+27
U+90\right) \hat\Zer_{,rrr}
   \bar\Zer
\end{eqnarray}
\begin{eqnarray}
\nonumber
&+&\frac{2}{r^3} \left(2 U^2+5
U-3\right)^2 \left(24138 U^9+399357 U^8+2535795
U^7+8866263 U^6
\right.\\\nonumber
&+&\left.20189321 U^5+31979265
U^4+34936825 U^3+20024625 U^2-4674375
U-9618750\right) \hat\Zer
   \bar\Zer_{,r}
\\\nonumber
&+&\frac{8}{r^2} \left(6 U^3+25 U^2+16
U-15\right)^2 \left(90 U^7+468 U^6+1763 U^5+5632
U^4+22704 U^3
\right.\\\nonumber
&-&\left.6480 U^2-35025 U+15750\right)
   \hat\Zer_{,r} \bar\Zer_{,r}
\\\nonumber
&+&\frac{2}{r} (U+3)^2
   \left(6 U^2+7 U-5\right)^3 \left(675 U^5+5741
U^4+15946 U^3+24570 U^2+10590 U-13950\right)
   \hat\Zer_{,rr} \bar\Zer_{,r}
\\\nonumber
&-&5 (U+3)^3 (11U-9) \left(6 U^2+7 U-5\right)^4 \hat\Zer_{,rrr}\bar\Zer_{,r}
\\\nonumber
&-&\frac{2}{r} (3 U+5)^2 \left(2 U^2+5
U-3\right)^3 \left(1377 U^5+11403 U^4+25994
U^3+19250 U^2-410 U-10350\right) \hat\Zer_{,r}
   \bar\Zer_{,rr}
\\\nonumber
&-&8 (1-2 U)^4 \left(3 U^2+14U+15\right)^3 \left(30 U^3+91 U^2+99 U+15\right)
   \hat\Zer_{,rr} \bar\Zer_{,rr}
\\\nonumber
&+&\frac{78}{r} (3 U+5)^2
   \left(2 U^2+5 U-3\right)^4 \left(3 U^3+15 U^2+25
U+50\right) \hat\Zer \bar\Zer_{,rrr}
\\\nonumber
&+&13 (3U+5)^3 (17 U-15) \left(2 U^2+5 U-3\right)^4
   \hat\Zer_{,r} \bar\Zer_{,rrr}
\\\nonumber
&-&\frac{8}{r^2} \left(3
U^2+14 U+15\right)^2 \left(198 U^7+1164 U^6+2627
U^5+17137 U^4+45972 U^3
\right.\\\nonumber
&+&\left.3140 U^2-38100
U+11475\right) \hat\Zer_{,t}
   \bar\Zer_{,t}
\\\nonumber
&-&\frac{2}{r} (U+3)^2 (3 U+5)^3
   \left(474 U^6+4575 U^5+13106 U^4+20156 U^3+1014
U^2-23685 U+8100\right) \hat\Zer_{,tr}
   \bar\Zer_{,t}
\\\nonumber
&+&5 (1-2 U)^2 (U+3)^3 (3 U+5)^4
   \left(4 U^2+23 U-9\right) \hat\Zer_{,trr}
   \bar\Zer_{,t}
\\\nonumber
&+&\frac{2}{r} (U+3)^3 (3 U+5)^2
   \left(2142 U^6+13005 U^5+21826 U^4+13968
U^3-4370 U^2-21115 U+8100\right) \hat\Zer_{,t}
   \bar\Zer_{,tr}
\\\nonumber
&+&24 \left(U^2-3 U-5\right)
   \left(6 U^3+25 U^2+16 U-15\right)^3
   \hat\Zer_{,tr} \bar\Zer_{,tr}
\\
&-&13 (1-2 U)^2 (U+3)^4
   (3 U+5)^3 \left(12 U^2+37 U-15\right)
   \hat\Zer_{,t} \bar\Zer_{,trr}
\Big\}.
\end{eqnarray}
\end{widetext}

\end{appendix}

\end{document}